\documentclass[12pt,tightenlines,eqsecnum,floats,aps,amsmath,amssymb,nofootinbib,prd,showpacs]{revtex4}

\usepackage{setspace} \usepackage{amsmath,amssymb,amsfonts,amsthm}
\usepackage{graphicx}

\newcommand{\be}{\begin{equation}}
\newcommand{\ee}{\end{equation}}
\newcommand{\beq}{\begin{eqnarray}}
\newcommand{\eeq}{\end{eqnarray}}

\begin{document}
\title{Cosmological consequences of the NonCommutative Geometry
  Spectral Action}
\author{Mairi Sakellariadou} 
\affiliation{King's College London, Department of Physics, Strand WC2R 
2LS, London, U.K.}

\begin{abstract}
\vspace{.2cm}
\noindent
Cosmological consequences of the noncommutative geometry spectral
action are presented. Neglecting the nonminimal coupling of the Higgs
field to the curvature, background cosmology remains unchanged, and
only the inhomogeneous perturbations will evolve differently from the
equivalent classical system. However, considering the nonminimal
coupling, corrections will be obtained even at the level of the
background cosmologies. Finally, the Higgs field may act as an
inflaton field, due to its nonminimal coupling with geometry.
\end{abstract}

\maketitle

\section{Introduction}
I will present cosmological consequences of the spectral action
functional in the noncommutative geometry model of particle
physics. The mathematical tools of noncommutative geometry~\cite{NCG}
offer the means to describe geometry and the geometric properties of
spaces, using the functions defined on the geometry, and the
properties of functions defined on the spaces, respectively.  

Much below the Planck scale, gravity can be safely considered as a
classical theory. However, as energies approach the Planck scale, the
quantum nature of space-time reveals itself and the Einstein-Hilbert
action becomes an approximation. At such energy scales, all forces,
including gravity, should be unified, so that all interactions
correspond to just one underlying symmetry. As a result, one may
expect that the nature of space-time would change at Planckian
energies, in such a way so that one can indeed recover the (familiar)
low energy picture of diffeomorphism, which governs General Relativity,
and internal gauge symmetries, which govern gauge groups on which the
Standard Model is based.

I will concentrate on a possible application of noncommutative
geometry, namely that space-time at the Planck length, $l_{\rm
  Pl}=\sqrt{G\hbar/c^3}$, is described by a different form of geometry
-- whose choice will certainly effect the phenomenological
consequences of the model -- than the continuous four-dimensional
manifold ${\cal M}$. I will work within a simple framework, based on
the hypothesis that space-time is the product of a continuous manifold
${\cal M}$ by a discrete space ${\cal F}$; the easiest generalisation
of a commutative space. Certainly, at Planckian energies the structure
of space-time must be noncommutative in a nontrivial way, so that its
low energy limit gives the simple product ${\cal M}\times{\cal F}$ we
are considering here.

I will focus on a model proposed by Connes and
collaborators~\cite{Chamseddine:2006ep}, aiming at describing the
Standard Model of particle physics coupled with gravity.  Following
the general noncommutative geometry approach, a physical system is
contained in the algebra of functions (space-time) represented as
operators on a Hilbert space (states), with the action and metric
properties encoded in a generalised Dirac operator.  The physical
Lagrangian is computed from the asymptotic expansion in the energy
scale $\Lambda$, of a natural action functionals, {\sl spectral
  action}, defined on noncommuatative spaces. The derived physical
Lagrangian is determined from the geometric input, namely the choice
of a finite dimensional algebra, thus the physical implications are
closely dependent on the underlying chosen geometry.

The successful outcome of the noncommutative geometry model proposed
by Connes and collaborators~\cite{Chamseddine:2006ep} lies in the fact
that the physical Lagrangian, obtained from the asymptotic expansion
of the spectral action, contains the full Standard Model Lagrangian
with additional Majorana mass terms for the right-handed neutrinos,
and gravitational and cosmological terms coupled to matter.  The
gravitational terms include: (i) the
Einstein-Hilbert action with a cosmological term, (ii) a topological
term which is related to the Euler characteristic of the space-time
manifold, (iii) a conformal gravity term having the Weyl curvature
tensor, and (iv) a conformal nonminimal coupling of the Higgs field
to gravity.  The existence of the two last contributions and the
dependence of the coefficients of the gravitational terms on the
Yukawa parameters of the particle physics content, diversify this
model from the usual minimal coupling of gravity to matter. Thus, one
expects some distinct consequences on early universe
cosmology~\cite{Nelson:2008uy}, and probably a more natural
inflationary mechanism~\cite{wm-infl}. It is important to note that
such features remain unique to this type of noncommutative geometry
models.

This noncommutative geometry spectral action offers, to my opinion, a
rich phenomenology (see, {\sl e.g.} Refs.~\cite{Nelson:2008uy},
~\cite{wm-infl}, ~\cite{mm-ep}), in both particle physics and
cosmology, which may provide tests for the theory, as well as some
insight for nontrivial noncommutative spaces near the Planck energy
scale.

%%%%%%%%%%%%%%%%%%%%%%%%%%%%

\section {Noncommutative geometry coupled to gravity}
Consider the simplest extension of the smooth four-dimensional
manifold, ${\cal M}$, by taking the product of it with a discrete
noncommuting manifold ${\cal F}$ of dimension $6$.  This internal
space has dimension $6$ to allow fermions to be simultaneously Weyl
and chiral (as within string theory), whilst it is discrete to avoid
the infinite tower of massive particles that are produced in string
theory.  The noncommutative nature of ${\cal F}$ is given by a
spectral triple, introduced by Connes~\cite{ac} as an extension of
the notion of Riemannian manifold to noncommutative geometry.  The
real spectral triple is given by the data $\left( {\cal A},{\cal H},
D\right)$, defined as follows: ${\cal A}$ is an involution of
operators on the Hilbert space ${\cal H}$, which is essentially the
algebra of coordinates. It is required the algebra to be unital, in
the sense that compact manifolds are considered.  $D$ is a linear
self-adjoint operator acting on ${\cal H}$, such that all commutators
$\left[ D,a \right]$ are bounded for $a\in{\cal A}$, that gives the
inverse line element.

By assuming that the algebra constructed in ${\cal M}\times {\cal F}$
is {\it symplectic-unitary}, the algebra ${\cal A}$ is restricted to
be of the form
\begin{equation}
\mathcal{A}=M_{a}(\mathbb{H})\oplus M_{k}(\mathbb{C})~,
\end{equation}
where $k=2a$ and $\mathbb{H}$ is the algebra of quaternions.  The
choice $k=4$ is the first value that produces the correct number of
fermions in each generation (note however that the number of
generations is an assumption in the theory), namely there are $k^2=16$
fermions in each of the three generations~\cite{Chamseddine:2007ia}.

The Dirac operator $D$ connects ${\cal M}$ and ${\cal F}$ via the
spectral action functional on the spectral triple.  It is defined as
${\rm Tr}\left( f\left(D/\Lambda\right)\right)$, where $f>0$ is a
cut-off (test) function and $\Lambda$ is the cut-off energy scale.
These three additional real parameters are physically related to the
coupling constants at unification, the gravitational constant and the
cosmological constant.  The asymptotic expression for the spectral
action, for large energy $\Lambda$, is of the form
\begin{equation}
{\rm Tr}\left(f\left({D\over\Lambda}\right)\right)\sim 
\sum_{k\in {\rm DimSp}} f_{k} 
\Lambda^k{\int\!\!\!\!\!\!-} |D|^{-k} + f(0) \zeta_D(0)+ {\cal O}(1)~,
\end{equation}
where $f_k= \int_0^\infty f(v) v^{k-1} {\rm d}v$ are the momenta of
the function $f$, the noncommutative integration is defined in terms
of residues of zeta functions, and the sum is over points in the
{\sl dimension spectrum} of the spectral triple.

The basic symmetry for a noncommutative space $\left( {\cal A},{\cal
  H}, D\right)$ is ${\rm Aut}({\cal A})$, which includes both the
diffeomorphism and internal symmetry transformations. The action
functional considered to obtain the physical Lagrangian has a bosonic
and a fermionic part. The bosonic action is a spectral function of the
Dirac operator, while the fermionic one has the simple linear form
$(\psi, D\psi)$, where $\psi$ are spinors defined on the Hilbert
space~\cite{sp-act}.  Applying this principle to the noncommutative
geometry of the Standard Model leads to the Standard Model action
coupled to Einstein and Weyl gravity plus higher order
nonrenormalisable interactions suppressed by powers of the inverse of
the mass scale of the theory.

To study the implications of this noncommutative approach coupled to
gravity for the cosmological models of the early universe, one can
concentrate just on the bosonic part of the action; the fermionic part
is however crucial for the particle physics phenomenology of the
model. Therefore, since for the time period we are concerned, namely
between the unification and the electroweak epoch, the Higgs field is
the most relevant one, we can just consider the part of the spectral
action given by ${\rm Tr}\left( f\left(D_A/\Lambda\right)\right)$,
where $D_A= D+A+\epsilon' JA J^{-1}$ is the Dirac operator with inner
fluctuations given by the gauge potentials of the form
$A=A^\dagger=\Sigma_k a_k [D,b_k]$, for elements $a_k,b_k\in {\cal
  A}$.

Writing the asymptotic expansion of the spectral action, a number of
geometric parameters appear, which describe the possible choices of
Dirac operators on the finite noncommutative space. These parameters
correspond to the Yukawa parameters of the particle physics model and
the Majorana terms for the right-handed neutrinos.  The Yukawa
parameters run with the renormalisation group equations of the
particle physics model. Since running towards lower energies, implies
that nonperturbative effects in the spectral action cannot be any
longer safely neglected, any results based on the asymptotic expansion
and on renormalisation group analysis can only hold for early
universe cosmology. For later times, one should instead consider the
full spectral action.

Finally, the empirical data taken as an input to the model are the
following ones: (i) there are 16 chiral fermions in each of the 3
generations, (ii) the photon is massless, and (iii) there are Majorana
mass terms for the neutrinos.

%%%%%%%%%

\subsection{Phenomenology}
The full Lagrangian of the Standard Model minimally coupled to
gravity, can be written~\cite{Chamseddine:2006ep} as the asymptotic
expansion of the spectral action on the product space-time ${\cal
  M}\times {\cal F}$.  One can then extract phenomenological
consequences. The relations between the gauge coupling constants have
been found to coincide with those obtained in Grand Unified
Theories. Namely, $ g_2^2=g_3^2=(5/3)g_1^2$, and $\sin^2 \theta_{\rm
  W}=(3/8)$, a value also obtained in SU(5) and SO(10).

The model has a number of successful outcomes. It leads to an
acceptable top quark mass of 179 GeV, neutrino mixing and see-saw
mechanism to give very light left-handed neutrinos are successfully
identified, and correct representations of the fermions with
respect to SU(3)$\times$SU(2)$\times$U(1) are
derived~\cite{Chamseddine:2006ep}.

Its drawbacks can be summarised as follows: (i) The unification of
gauge couplings with each other and Newton constant do not meet at one
point. (ii) The mass of the Higgs field is of the order of 170 GeV, a
value which is recently ruled out experimentally. However, one should
always keep in mind that higher order contributions to the Higgs
potential may modify the prediction for the Higgs mass. (iii) No new
particles besides those of the Standard Model are predicted and this
may be problematic if new physics is discovered at the Large Hadron
Collider. (iv) There is no explanation of the number of
generations. (v) There are no constraints on the values of Yukawa
couplings.

Before however drawing any {\sl unfair} conclusions, one must keep in
mind that we have considered the simplest generalisation of the
commutative geometry, which certainly is an effective theory. As I
have stated previously, at Planckian energies nontrivial
generalisations of noncommutative spaces must be considered, which
will certainly change the particle spectrum. In other words, before
drawing any conclusions, one should firstly, include higher
order corrections to the spectral action, to show gauge couplings
unification, and thus get an accurate prediction for the Higgs mass,
and secondly, find the noncommutative space whose limit is
the product ${\cal M}\times{\cal F}$.

%%%%%%%%%%%%%%%
\subsection{Cosmological consequences}
In what follows, I will briefly summarise the results we have obtained
on some cosmological consequences of this noncommutative geometry
spectral action. In particular, I will first discuss corrections to
Einstein's equations and then I will present a possible inflationary
mechanism through the Higgs field.
\subsubsection{Corrections to Einstein's equations}
For the purpose of our work, namely extracting cosmological
consequences of the noncommutative geometry coupled to gravity
approach, we are only interested in the gravitational part of the
action:
\be\label{eq:action1} 
{\cal S}_{\rm grav} = \int \left( \frac{1}{2\kappa_0^2} R + \alpha_0
C_{\mu\nu\rho\sigma}C^{\mu\nu\rho\sigma} + \tau_0 R^\star R^\star -
\xi_0 R|{\bf H}|^2 \right) \sqrt{g} {\rm d}^4 x~, 
\ee
where $ {\bf H} $ is a rescaling ${\bf H} = (\sqrt{a f_0}/\pi) \phi $
of the Higgs field $\phi$ to normalise the kinetic energy and
$f_0=f(0)$. The action of quaternions $\mathbb{H}$ can be represented
in terms of Pauli matrices $\sigma^\alpha$, namely $q=f_0 + \sum i
f_\alpha \sigma^\alpha$, with $a$
\be
 a={\rm Tr} \left( Y^\star_{\left(\uparrow 1\right)}
Y_{\left(\uparrow 1\right)} +  Y^\star_{\left(\downarrow 1\right)}
Y_{\left(\downarrow 1\right)}
+  3\left( 
 Y^\star_{\left(\uparrow 3\right)} Y_{\left(\uparrow 3\right)}
+  Y^\star_{\left(\downarrow 3\right)} Y_{\left(\downarrow 3\right)}
\right)\right)~;
\ee
the $Y$'s are used to classify the action of the Dirac operator
and give the fermion and lepton masses, as well as lepton mixing, in
the asymptotic version of the spectral action. 
The coupling constants
in Eq.~(\ref{eq:action1}) are
\beq \frac{1}{\kappa_0^2} = \frac{ 96 f_2 \Lambda^2 - f_0
  c^2}{12\pi^2}~,\nonumber\\ \alpha_0 = -\frac{3f_0}{10\pi^2}~~,
~~\tau_0 = \frac{11f_0}{60\pi^2}~~,~~ \xi_0 = \frac{1}{12}~; \eeq
$\Lambda$ is the renormalisation cut-off and $c$ is expressed in
terms of $Y_R$ which gives the Majorana mass matrix, $c={\rm Tr}
\left( Y^\star_R Y_R \right)$. 

The first two terms in Eq.~(\ref{eq:action1}) give the Riemannian
curvature with a contribution from the Weyl curvature; the second term
is the action for conformal gravity.  The third is a topological term
integrating to the Euler characteristic of the manifold,
\be 
R^\star R^\star =\frac{1}{4} \epsilon^{\mu\nu\rho\sigma}
\epsilon_{\alpha\beta\gamma\delta} R^{\alpha\beta}_{\mu\nu}
R^{\gamma\delta}_{\rho\sigma} ~, 
\ee
hence nondynamical. The fourth is the scalar mass term.

The equations of motion arising from Eq.~(\ref{eq:action1})
read~\cite{mannheim}
\be\label{eq:EoM1} 
R^{\mu\nu} - \frac{1}{2}g^{\mu\nu} R -
\alpha_0\kappa_0^2\delta\left(\Lambda\right) \left[
  2C^{\mu\lambda\nu\kappa}_{\ \ \ \ \ \ ;\lambda ; \kappa} -
  C^{\mu\lambda\nu\kappa}R_{\lambda \kappa}\right]
 = 
 \kappa_0^2\delta\left(\Lambda\right)T^{\mu\nu}_{\rm matter}~, 
\ee
\be \mbox{where} ~~~~~~\delta\left( \Lambda\right) \equiv [1
  -2\kappa_0^2 \xi_0|{\bf H}|^2]^{-1}~. \nonumber 
\ee

Neglecting the nonminimal coupling between the geometry and the Higgs
field, i.e. setting $\phi = 0$, we have explicitly
shown~\cite{Nelson:2008uy} that noncommutative corrections to
Einstein's equations are present only for inhomogeneous and
anisotropic space-times.  This is the effect of the purely geometrical
terms; the term $R^\star R^\star$ is topological and hence plays no
r\^ole in dynamics, while the term
$C_{\mu\nu\rho\sigma}C^{\mu\nu\rho\sigma}$ vanishes for homogeneous
and isotropic metrics. As a result, one is left with just the
Einstein-Hilbert term.  One could have already anticipated this
result, since any effects on the noncommutativity of space-time
coordinates should vanish for homogeneous and isotropic backgrounds,
having all their points equivalent.

Considering the nonminimal coupling of the Higgs field to the
curvature, which cannot be neglected as we approach the early universe
era, we have however obtained~\cite{Nelson:2008uy} corrections even
for background cosmologies. This may lead to important
phenomenological consequences~\cite{Nelson:2008uy}, since the effect
of the nonminimal coupling of the Higgs field to the curvature results
to a larger Higgs mass, within the context of static geometries.

%%%%%%%%%%%%%%%%%%%%%%%%
\subsubsection{Higgs field driven inflation}
The gravitational and Higgs part of the asymptotic expansion of the
spectral action read~\cite{Chamseddine:2006ep} \be
\label{eq:action2} 
{\cal S}_{\rm grav} = \int \left( \frac{1}{2\kappa_0^2} R + \alpha_0
C_{\mu\nu\rho\sigma}C^{\mu\nu\rho\sigma} + \tau_0 R^\star
R^\star+\gamma_0 -\xi_0 R|{\bf H}|^2
+\frac{1}{2} |D_\mu {\bf H}|^2 + V\left( |{\bf H}|\right) \right)
\sqrt{g} {\rm d}^4 x~, 
\ee
where the potential $V\left(|{\bf H}|\right)= \lambda_0|{\bf H}|^4 -
\mu_0^2 |{\bf H}|^2$ is the standard Higgs potential, and the
$\kappa_0^2, \alpha_0,\tau_0,\lambda_0,\mu_0$ are specified in terms
of the cut-off energy scale $\Lambda$, the couplings and the
coefficients $f_k$.
 
The above action, Eq.~(\ref{eq:action2}), implies that, in addition to
the cosmological constant term $\gamma_0$, which we will neglect here,
the geometry is nonminimally coupled to the Higgs field. Such
modification to the Einstein-Hilbert action is indeed the one required
so that the amplitude of the perturbations during Higgs field
inflation gets reduced, allowing the Higgs field to satisfy
simultaneously the Standard Model requirements, as well as those
imposed to inflation from the WMAP5 measurements of the Cosmic
Microwave Background temperature anisotropies.

Let me be more specific: It is a long standing proposal that the
scalar field of the Standard Model, namely the Higgs field, could play
the r\^ole of the inflaton.  However, within the general relativistic
cosmology, in order to get the correct amplitude of density
perturbations, the Higgs mass would have to be some 11 orders of
magnitude higher that the one required by particle physics.  Luckily,
this requirement does not apply in the context of the noncommutative
geometry coupled to gravity approach.

To reduce the amplitude of the induced Higgs field perturbations, from
$\lambda_0$ to $\lambda_0/\xi^2_0$, a nonminimal coupling, like the
one that naturally appears in the context of the noncommutative
geometry, has been postulated~\cite{Bezrukov:2007ep}. Thus, the Higgs
field could play the r\^ole of the inflaton, provided inflation took
place at a scale higher than the strong-weak unification scheme at
$10^{17}$GeV.  Analysis of the running of couplings with the cut-off
scale would determine the energy scale of inflation, a still open
issue since it has been only studied neglecting the nonminimal
coupling between the Higgs field and the curvature, which is indeed
vital here. 

Let me now emphasise that the Higgs driven inflation presented within
this approach, does not suffer from the criticism against the {\sl
  conventional} Higgs field driven inflationary models.  It has been
recently argued~\cite{Burgess:2009ea} that corrections to the
semi-classical approximation, which are typically neglected in {\sl
  standard} inflationary models, may no longer be neglected for such
{\sl exotic} (Higgs field driven) scenarios. This is certainly a valid
criticism, which luckily does not apply in the noncommutative
geometry context.  The reason is simple: In {\sl conventional} Higgs
inflation there is a strong coupling, namely $\xi_0\sim 10^4$, between
the Higgs field and the Ricci curvature scalar. Thus, the effective
theory ceases to be valid beyond a cut-off scale $m_{\rm Pl}/\xi_0$.
However, one should know the Higgs potential profile for the field
values relevant for inflation, namely $m_{\rm Pl}/\sqrt\xi_0$, and
such values are much higher than the cut-off of the validity of the
effective theory. This argument is clearly not applicable to the
noncommutative Higgs driven inflation, where $\xi_0=1/12$.

%%%%%%%%%%%%%%%%%%%%%%%
\section{Conclusions}
I have presented some cosmological consequences arising from the
asymptotic expansion of the spectral action functional in the
noncommutative geometry model of particle physics. This proposal has
the potential of offering a concrete and fundamental theoretical
context to build a cosmological model.

I have first discussed that neglecting the nonminimal coupling of the
Higgs field to the curvature, noncommutative geometry corrections to
Einstein's equations are present only for inhomogeneous and
anisotropic cosmologies. However, considering the nonminimal coupling,
there are corrections even for background cosmologies.

I have then presented how natural inflation may indeed occur as a
consequence of the nonminimal coupling between geometry and the Higgs
field. The term which has been introduced {\sl ad hoc} in order to
achieve a Higgs field driven inflation, arises naturally in the
context of the noncommutative geometry coupled to gravity approach.

Certainly, many interesting questions, concerning both cosmological
consequences as well as particle physics implications, remain to be
addressed and  we plan to inevstigate them in a future study.

%%%%%%%%%%%%%%%%%%%%%%%%%%
\section{Acknowledgements}
It is a pleasure to thank the organisers of the interesting and
stimulating meeting {\sl XXV Max Born Symposium}, in Wroclaw (Poland)
for inviting me to participate and present these results.  This work
is partially supported by the European Union through the Marie Curie
Research and Training Network {\sl UniverseNet} (MRTN-CT-2006-035863).

\end{document}